\documentclass[11pt]{article}
\usepackage{moriond,epsfig}

\def\be{\begin{equation}}
\def\ee{\end{equation}}
\def\bea{\begin{eqnarray}}
\def\eea{\end{eqnarray}}

\begin{document}
\vspace*{4cm}
\title{Cosmological Markov Chain Monte Carlo simulation with cmbeasy}

\author{ Christian M. M{\"{u}}ller}

\address{Institut f{\"{u}}r Theoretische Physik, Philosophenweg 16, 69120 Heidelberg, Germany}

\maketitle\abstracts{ We introduce a Markov Chain Monte Carlo simulation and data analysis
  package for the cosmological computation package {\sc Cmbeasy}.  We have taken
  special care in implementing an adaptive step algorithm for the
  Markov Chain Monte Carlo in order to improve convergence.
  Data analysis routines are provided which allow to
  test models of the Universe against up-to-date measurements of the Cosmic
  Microwave Background, Supernovae Ia and Large Scale Structure. The observational data
is provided with the software for convenient usage.
  The package is publicly available as  part of the {\sc Cmbeasy} software at www.cmbeasy.org.
}

\newcommand{\class}[1]{{\tt #1}}
\newcommand{\postd}{\pi(\theta|X)}
\newcommand{\covM}{\textbf{S}}
%%%%%%%%%%%%%%%%%%%%%%%%%%%%%%%%%%%%%%%%%%%%%%%%%%%%%%%%%%%%%%%%%%%%%%%%%%%%%%%
\section{Introduction}
The wealth of recent  precision measurements in cosmology 
\cite{Bennett:2003bz,Hinshaw:2003ex,Kogut:2003et,Kuo:2002ua,Readhead:2004gy,Dickinson:2004yr,Riess:2004nr,Tonry:2003zg,Knop:2003iy,Barris:2003dq,Percival:2001hw,Verde:2002,Peacock:2001gs,Tegmark:2003ud,Tegmark:2003uf} 
places stringent constraints on any model of the Universe. Typically, such a model is given in terms of a number
of cosmological parameters. Numerical tools, such as {\sc Cmbfast} \cite{Seljak:1996is}, 
\textsc{Camb}  \cite{Lewis:1999bs} and \mbox{\sc Cmbeasy}  \cite{Doran:2003sy}, permit to calculate 
the prediction of a given model  for the observational data. While these tools are  comparatively fast, 
scanning the parameter space for the most likely model and confidence regions can become a matter of 
time and computing power. 
The cost of evaluating models on a n-dimensional grid in parameter space 
increases exponentially with the number of parameters. 
In contrast, the Markov Chain Monte Carlo (MCMC) method scales 
roughly linearly with the number of parameters \cite{Christensen:2000ji,Christensen:2001gj,Lewis:2002ah}.  The MCMC
method has already been used to constrain  various models \cite{Knox:2001fz,Caldwell:2003hz,Kosowsky:2002zt,Verde:2003ey}.
A popular tool for setting up MCMC simulations is the \textsc{Cosmo-Mc} package \cite{Lewis:2002ah} 
for the  \textsc{Camb} code, an improved proposal distribution for the local Metropolis algorithm 
has been proposed in \cite{Slosar:2003da}.

\begin{figure}
\begin{center}
\includegraphics[scale=7]{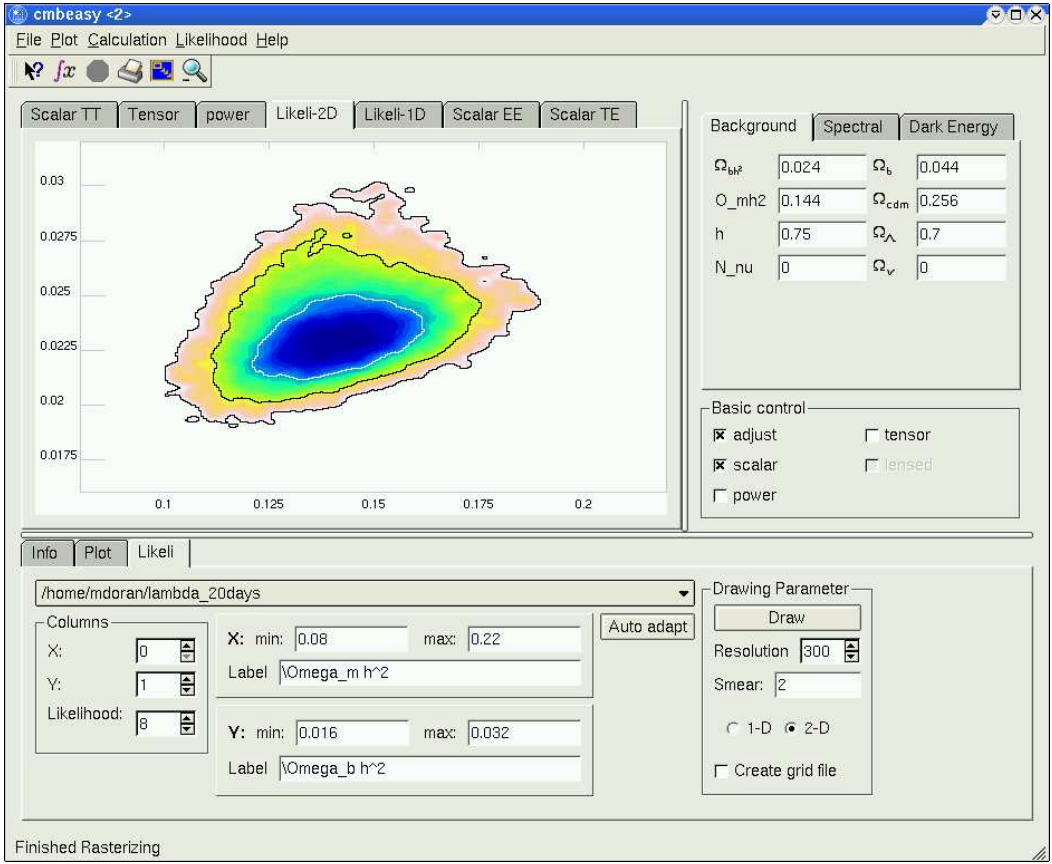}
\caption{\label{fig::gui}The graphical user interface of {\sc Cmbeasy}. It can be used to 
marginalize, visualize and print the one and two dimensional likelihoods from the MCMC chains. 
Shown is the marginalized likelihood in the $\Omega_m h^2 - \Omega_b h^2$ plane of a $\Lambda$CDM model. } 
\end{center}
\end{figure}

We  introduce the \class{AnalyzeThis} package\footnote{It is part of the cmbeasy v2.0 release.}
 \cite{Doran:2003ua} for {\sc Cmbeasy}. It includes a
parallel MCMC driver, as well as
routines 
to calculate the likelihood of a model with respect to various data sets. 
We took special care in designing a step-proposal strategy that leads to fast
convergence and mixing of the chains. The automated step optimization 
considerably improves performance and is rather convenient. 
The raw data files can be processed from within a graphical user interface (gui).
Using the gui, one can marginalize, visualize and print  one and two
dimensional likelihood surfaces (see figure \ref{fig::gui}).

%%%%%%%%%%%%%%%%%%%%%%%%%%%%%%%%%%%%%%%%%%%%%%%%%%%%%%%%%%%%%%%%%%%%%%%%%%%%%%%%%%%%%%%%%%%%%%%%%%%%
\section{Markov Chain Monte Carlo simulation}\label{sec::simulation}
%%%%%%%%%%%%%%%%%%%%%%%%%%%%%%%%%%%%%%%%%%%%%%%%%%%%%%%%%%%%%%%%%%%%%%%%%%%%%%%%%%%%%%%%%%%%%%%%%%%%%%%%%
In the following, we will assume the reader is familiar with the basic ideas of Markov Chain
Monte Carlo simulation  \cite{Gilks} and the Metropolis algorithm\cite{Metropolis:am} . We will now describe our 
implementation as contained in the released version \cite{Doran:2003ua}. 

The global Metropolis algorithm chooses new steps for a Markov Chain via a symmetric proposal distribution 
$q(\theta_{i-1},\theta_i)=q(\theta_{i},\theta_{i-1})$, where $\theta_i$ is a parameter vector that specifies a given model. In our implementation, we assume flat priors $P(\theta)$ on the parameters, and we assign
likelihood zero to any parameter set that has at least one point outside the prior. 

A very important aspect of MCMC is to test when the chains are sampling from the underlying distribution.
Since at the beginning, the chain migrates from its random
starting point to regions of higher likelihood, there is a ``burn-in'' associated with each chain 
that must be eleminated prior to parameter estimation. In principle,
it may be difficult to tell from a single chain if it has converged. In MCMC, one therefore uses several
chains with random starting points and monitors mixing and
convergence. Our implementation employs the convergence test of Gelman and Rubin\cite{Gelman_Rubin}. The key ingredient for
this test is a parameter $R$ which can be computed from previous chain points. This parameter is 
a comparison of the variance within the chains compared to the variance between different chains. A value of
$R < 1.2$ for each parameter indicates the chains have converged and all previous points should be removed.  
 If one uses the gui for chain analysis, the burn-in is automatically removed. 

Since there is no generally accepted procedure to determine when one has generated enough chain points for 
reliable estimation, the algorithm just runs indefinetly in our implementation. 
However, any ``breaking-criterion''\cite{Dunkley:2004sv} may be implemented easily. The chains may be monitored with external programs during the run.

The number of steps needed for good convergence and mixing depends
strongly on the step proposal distribution. If the proposed steps are too
large, the algorithm will frequently reject steps, giving slow
convergence of the chain. If, on the other hand, the proposed steps
are too small, it will take a long time for the chain to explore the
likelihood surface, resulting in slow mixing.  In the ideal case the
proposal distribution should be as close to the posterior distribution  as possible -- which
unfortunately is not known a priori.
While a simple Gaussian proposal distribution with step sizes $\sigma_k$ is
sufficient, it is not optimal in terms of computing costs if cosmological parameters
are degenerate.
%%%%%%%%%%%%%%%%%%%%%%%%%%%%%%%%%%%%%%%%%%%%%%%%%%%%%%%%%%%%%%%%%
Instead of using a naive Gaussian proposal distribution, we sample from a multivariate
Gaussian distribution with covariance matrix estimated from the previous points in the chains. 
By taking into account the covariances
among the parameters, we effectively approximate the likelihood contour in extent and orientation --  the
Gaussian samples are taken along the principal axis of the likelihood contour.

The convergence can be further improved by scaling the covariance
matrix with a variable factor $\alpha$. Using $\alpha$, we
can cope better in situations where the projected likelihood takes on
banana shapes such as in \cite{Kosowsky:2002zt}.  It also improves the
convergence during the early stages when the low number of points
available limits the estimate of the covariance matrix.  We dynamically
increase $\alpha$ if a chain takes steps too often, while we decrease
$\alpha$ if the acceptance rate is too low.
By this procedure the convergence is speeded up by a factor of about four compared to naive Gaussian sampling.

One can show that modifying the proposal distribution  based on previous chain data during
the run may lead to  a wrong stationary distribution.
Therefore, we only apply the  dynamical strategy of finding an optimal step proposal during
the early stages of the simulation. When the convergence
is better than $R=1.2$ and the chain has calculated a certain number of points,
we freeze in  the step proposal distribution.

%%%%%%%%%%%%%%%%%%%%%%%%%%%%%%%%%%%%%%%%%%%%%%%%%%%%%%%%%%%%%%%%%%%%%%%%%%%%%%%%%%%%%%%%%%%%%%%%%%%
\section{The Software}\label{sec::user}

The package is  part of {\sc Cmbeasy} and 
consists of two main components. The first one is a
MCMC driver using LAM/MPI for parallel execution of  each
chain. The second one is the \class{AnalyzeThis} class which is designed to
evaluate the likelihood of a given model with respect to various data
sets. These sets include the latest data of WMAP TT and TE \cite{Hinshaw:2003ex,Kogut:2003et},
ACBAR \cite{Kuo:2002ua}, CBI \cite{Readhead:2004gy}, VSA \cite{Dickinson:2004yr} , 2dFGRS
\cite{Percival:2001hw,Verde:2002,Peacock:2001gs}, SDSS \cite{Tegmark:2003ud,Tegmark:2003uf},
the SNe Ia
compilations of Riess et al. \cite{Riess:2004nr}, Tonry et al.~\cite{Tonry:2003zg} and Knop et. al.
\cite{Knop:2003iy} as well as  the IfA Deep Survey SNe Ia data \cite{Barris:2003dq}.  
Data files for all experiments are included for
convenience. New data sets are added continuously to the code. 

The example MCMC driver consists of two routines: master() and
slave(). Using LAM/MPI for parallel computing, one master and up to
ten slaves may be started. 
The master() will determine the initial random starting position for
each chain. In a never ending  loop, it then sends the parameters
to the slave()'s and collects the results when the computation is finished.
Whenever a step has been successful, it stores the parameters and
likelihoods of the last step together with  the number of  times the chain remained 
at the same point in parameter space in a new line of one file per chain.
The master() monitors convergence and mixing and determines
the next step for the slave(). Before freeze in, the covariance is 
estimated and the step proposal accordingly modified. After freeze in,
the proposal distribution remains unchanged.

The gui may be used to process the raw output files of the MCMC chains. It enables quick and convenient 
analysis of MCMC simulations.
To get started immediately, we include 
raw data from an example MCMC run  in the resources directory of cmbeasy.
Two and one dimensional marginalized likelihoods may then be plotted and printed from within
the gui (see figure \ref{fig::gui}).

%%%%%%%%%%%%%%%%%%%%%%%%%%%%%%%%%%%%%%%%%%%%%%%%%%%%%%%%%%%%%%%%%%%%%%%%%%%%%%%%%%%%%%%
\section{Conclusions}\label{sec::conclusion}
We have introduced the {\sc AnalyzeThis} package, which can be used
to constrain cosmological models using observational data sets.
The \class{AnalyzeThis}  class provides many functions to compute the likelihood of a model with respect
to  measurements of the CMB, SNe Ia and Large Scale Structure.

In order to constrain models of the Universe with a substantial number
of parameters, we include a Markov Chain Monte Carlo driver. 
As the MCMC step strategy 
determines the convergence speed of the chains, we implemented a multivariate Gaussian sampler
with an additional dynamical scaling.

\bibliographystyle{unsrt}

\end{document}